\documentclass{article}

\usepackage[utf8]{inputenc}
\usepackage{amsmath,amsthm,amssymb,amsfonts}
\usepackage{graphicx}
\usepackage{listings}
\usepackage{xcolor}
\usepackage{hyperref}

\usepackage{inconsolata}
\definecolor{forestgreen(web)}{rgb}{0.13, 0.55, 0.13}
\definecolor{fireenginered}{rgb}{0.81, 0.09, 0.13}
\newcommand{\dif}[1]{\ensuremath{\,\text{d}#1}}
\newcommand{\df}[2]{\ensuremath{{\text{d}#1}/{\text{d}#2}}}
\newcommand{\eq}[1]{eq.~\ref{#1}}
\newcommand{\citeeq}[2]{eq.~{#1} in ref.~\cite{#2}}
\newcommand{\code}{\lstinline}
\let\oldcite\cite
\renewcommand{\cite}{\unskip~\oldcite}

\newcommand{\reffig}[1]{fig.~\ref{#1}}
\newcommand{\refsec}[1]{sec.~\ref{#1}}
\newcommand{\refcite}[1]{ref.~\cite{#1}}

\title{A fast C++ implementation of thermal functions}
\author{Andrew Fowlie}
\date{}

\begin{document}

\maketitle

\begin{abstract}
We provide a small C++ library with Mathematica and Python interfaces for computing thermal functions, defined
\begin{equation*}
J_\text{B/F}(y^2) \equiv \Re \int_0^\infty x^2 \log\left[1 \mp e^{-\sqrt{x^2 + y^2}} \right] \dif{x},
\end{equation*}
which appear in finite-temperature quantum field theory and play a role in phase-transitions in the early Universe, including baryogenesis, electroweak symmetry breaking and the Higgs mechanism. The code is available from \url{https://github.com/andrewfowlie/thermal_funcs}.
\end{abstract}

\section*{PROGRAM SUMMARY}
\begin{description}
\item [Program title] \code{thermal_funcs}
\item [Licensing provisions] BSD 3-Clause
\item [Programming language] C++, C interface to Mathematica and SWIG interface to Python
\item [Nature of problem] Thermal functions appear in finite-temperature quantum field theory and influence phase transitions in the early Universe. They have no closed-form solution. Studying phase transitions requires repeated evaluations of, inter alia, thermal functions. We provide several fast and accurate methods, and first and second derivatives.
\item [Solution method] We implement numerical quadrature, a Bessel function representation, asymptotic solutions expressed in terms of zeta functions and polylogarithms, and approximations and limits.
\end{description}

\section{Introduction}

The history and fate of our Universe are influenced by phase transitions between vacuum states of a quantum field theory. Transitions between vacua alter the symmetries and particles that we observe in Nature. The celebrated electroweak phase transition
generated mass (see e.g., \refcite{Dawson:1998yi}) and a phase transition in the early Universe (or indeed the aforementioned electroweak phase transition) may have been a critical ingredient in baryogenesis (see e.g., \refcite{White:2016nbo}).
The presence of a phase transition in a theory and the order of the transition --- whether it is smooth or violent with a discontinuous change in free energy --- are governed by the temperature dependence of the free-energy. 

To make numerical studies of phase transitions in finite-temperature quantum field theory, we must evaluate the thermal functions, which encode the temperature dependence of the free-energy (see e.g., \citeeq{2.12}{Curtin:2016urg}):
\begin{equation}\label{eq:def}
J_\text{B/F}(y^2) \equiv \Re \int_0^\infty x^2 \log\left[1 \mp e^{-\sqrt{x^2 + y^2}} \right] \dif{x},
\end{equation}
where $y^2$ is real. There exists no closed-form solution. Note that the thermal functions are occasionally defined with an additional plus-minus sign or written as functions of $y$. See e.g., \citeeq{2.16}{Wainwright:2013maa},
\begin{equation}
I_{\pm}(y) \equiv \pm \Re \int_0^\infty x^2 \log\left[1 \mp e^{-\sqrt{x^2 + y^2}} \right] \dif{x}.
\end{equation}
With this convention $y$ may be complex, whereas $y^2$ must be real. For easier interfaces, we favour writing the thermal functions as functions of a real argument, $y^2$. In physical applications, the argument $y^2$ is a field-dependent mass squared divided by temperature squared, $m^2(\phi) / T^2$. The mass squared may be tachyonic, i.e., negative.

The functions appear in one-loop corrections to the free-energy of scalar fields in finite-temperature quantum field theory. Since they impact the free-energy, the functions play a role in phase transitions between vacuum states and thus the phase history of our Universe. Investigating phase transitions requires us to trace the vacua of the free-energy as the Universe cools; this requires, inter alia, many evaluations of the thermal functions. To ease such calculations, and to investigate their behaviour and that of different methods for their evaluation, we present a new C++ code for thermal functions and their first and second derivatives. The functions were previously implemented in Python in \code{CosmoTransitions}\cite{Wainwright:2011kj}.

\section{Numerical methods}\label{sec:methods}

We evaluate the integrals in \eq{eq:def} with several methods; this validates our numerical results and demonstrates the precision and speed of approximations:

\begin{enumerate}
    \item \code{quad}: Numerical quadrature, treating singularities in the integrand via \code{QUADPACK} algorithms in \code{GSL}.
    \item \code{bessel}: A Bessel function representation recognised in \refcite{PhysRevD.45.2685}, extended to $y^2 < 0$.
    \item \code{taylor}: A Taylor expansion of the integral in $y$.
    \item \code{zeta}: Asymptotic formulae for the integrals in terms of Hurwitz zeta functions and polylogarithms, including an Euler-Maclaurin summation of the Hurwitz zeta function, which is absent in common \code{GSL} and \code{boost} libraries.
    \item \code{approx}: Leading-order approximations for the integrals.
    \item \code{lim}: Asymptotic bounds for the integrals if $y^2 <0 $.
\end{enumerate}
Each of these requires care, detailed in the following sections.

\subsection{Quadrature (\normalfont\code{quad})}

For $x^2 + y^2<0$, we rewrite the integrand as
\begin{equation}
x^2 \Re \log\left[1 \mp e^{-\sqrt{x^2 + y^2}}\right] = 
\frac 12 x^2 \left[\log 2 + \log \left(1 \mp \cos \sqrt{|x^2 + y^2|}\right)\right].
\end{equation}
We use a special \code{gsl_log1p} function for precise evaluations of $\log(1 + x)$.

Special care must be taken about the singularities in the integrand if $y^2\le0$. They occur whenever
\begin{equation}
\sqrt{|x^2 + y^2|} = n\pi, 
\end{equation}
for even integers $n = 0, 2, \cdots$ for $J_\text{B}$ and for odd integers $n = 1, 3, \cdots$ for $J_\text{F}$. The singularities are enumerated and passed as known singular points to 
\code{gsl_integration_qagp} in the method \code{quad}. There is no ambiguity in the phase of $\sqrt{|x^2 + y^2|}$ since $\cos x$ is even.

\subsection{Bessel function representation (\normalfont\code{bessel}, \normalfont\code{approx}, \normalfont\code{zeta} and \normalfont\code{lim})}

The thermal functions may be represented by an infinite sum of modified Bessel functions of the second kind (see e.g., \citeeq{2.13}{Curtin:2016urg}):
\begin{equation}\label{eq:bessel}
J_\text{B/F}(y^2) = - y^2  \sum_{n=1}^{\infty} \frac{(\pm1)^n}{n^2} \Re K_2(n y).
\end{equation}
We implement this infinite sum explicitly in \code{bessel}, breaking once it converges to within a specified absolute and relative tolerance, or once a maximum number of terms is exceeded. There is no ambiguity in the phase of $\sqrt{y^2}$ since the real part of the $K_2$ Bessel function is even.

For $y^2 \gg 0$, the $K_2$ Bessel functions decay faster than exponentially,
\begin{equation}\label{eq:K2}
K_2(x) \approx \sqrt{\frac{\pi}{2 x}}e^{-x},
\end{equation}
such that the leading term,
\begin{equation}
J_\text{B/F}(y^2) \sim \mp \sqrt{\frac{\pi}{2}} y^{3/2} e^{-y},
\end{equation}
is a reasonable approximation to the thermal functions, which we implement in \code{approx}. The sum in \eq{eq:bessel} with the approximation in \eq{eq:K2} may be identified with a polylogarithm, such that the thermal functions are
\begin{equation}\label{eq:poly}
J_\text{B/F}(y^2) \approx - \sqrt{\frac{\pi}{2}} |y|^{3/2} \, \text{Li}_{5/2}(\pm e^{-y}),
\end{equation}
where $\text{Li}_s(x)$ is a polylogarithm function (forthcoming). This is implemented in \code{zeta}.

For $y^2 < 0$, we must evaluate Bessel functions for complex arguments. In \code{bessel}, we utilise the identity between modified and unmodified Bessel functions of the second kind,
\begin{equation}\label{eq:real}
\Re K_2(x i) = \frac{\pi}{2} Y_2(x).
\end{equation}
For $x \gg 0$, there is an asymptotic approximation for the Bessel functions of the second kind (see \citeeq{14}{K2}),
\begin{equation}
Y_2(x) \sim \sqrt\frac{2}{\pi x} \sin(x - 5/4\pi) = 
\sqrt\frac{1}{\pi x} [\cos x - \sin x].
\end{equation}
Thus we require the sum
\begin{equation}
\sum_{n=1}^{\infty} \frac{(\pm1)^n}{n^{5/2}} \left[\cos(nx) - \sin(nx)\right].\\
\end{equation}
This sum is related to polylogarithms and from an $s=5/2$ case of Hurwitz's formula\cite{Hurwitz}, we obtain
\begin{equation}\label{eq:hurwitz}
\sum_{n=1}^{\infty} \frac{(\pm1)^n}{n^{5/2}} \left[\cos(nx) - \sin(nx)\right] =
\begin{cases} 
    -\frac{16 \pi^2}{3} \zeta\left(-\frac32, -\frac{x}{2\pi}\right)\\
    -\frac{16 \pi^2}{3} \zeta\left(-\frac32, \frac12 - \frac{x}{2\pi}\right)
\end{cases}
\end{equation}
which are valid for $-2\pi \le x \le 0$ and $-\pi \le x \le \pi$, respectively. The argument $x$ must be shifted by $2\pi$ until it lies inside the appropriate interval. The function $\zeta(s, a)$ is a Hurwitz zeta function (forthcoming). 

With these results, we express the thermal functions in terms of Hurwitz zeta functions,
\begin{equation}
J_\text{B/F}(y^2) \approx
\begin{cases} 
  - |y|^{3/2} \frac{8 \pi^{5/2}}{3\sqrt{2}} \zeta\left(-\frac32, -\frac{\Im y}{2\pi}\right)\\
  - |y|^{3/2} \frac{8 \pi^{5/2}}{3\sqrt{2}} \zeta\left(-\frac32, \frac12 - \frac{\Im y}{2\pi}\right),
\end{cases}
\end{equation}
where $\Im y$ denotes the imaginary part of $y$. This is implemented in \code{zeta}.

We can bound the thermal functions by noting that
\begin{equation}
|J_\text{B/F}(y^2)| \sim \left|\sum_{n=1}^{\infty} \frac{(\pm1)^n}{n^{5/2}} |y|^{3/2} \sqrt{\frac{\pi}{2}} \sin(n \Im y - 5/4\pi) \right| \le \zeta (5/2) \sqrt{\frac{\pi}{2}} |y|^{3/2}.
\end{equation}
The extrema of the Hurwitz zeta function in \eq{eq:hurwitz}, however, lead to a sharper bound:
\begin{equation}\label{eq:bound}
{-}0.024 \cdot\frac{8\pi^{5/2}}{3} |y|^{3/2} \le J_\text{B/F}(y^2) \le 0.032 \cdot\frac{8\pi^{5/2}}{3} |y|^{3/2}. 
\end{equation}
The numerical factors originate from the extrema of the Hurwitz zeta function $\zeta(-3/2, a)$.  We implement this limit numerically in \code{lim}. Finally note that the leading term (implemented in \code{approx}),
\begin{equation}
J_\text{B/F}(y^2) \sim \pm |y|^{3/2} \sqrt{\frac{\pi}{2}} \sin(\Im y - \pi/4), 
\end{equation}
may be a reasonable approximation to the sum.

\subsection{Hurwitz zeta function and polylogarithm}

The Hurwitz zeta function is defined by
\begin{equation}
\zeta(s, a) = \sum_{n=0}^\infty \frac1{(a + n)^s},
\end{equation}
and its analytic continuation for $\Re s \le 1$. Since a general Hurwitz zeta function is absent in \code{boost} and \code{GSL}, though present in e.g., the older \code{CEPHES} library, we implement this function via a Euler-Maclaurin summation described in e.g., \refcite{Johansson2015}. This involves writing the function as
\begin{equation}
\zeta(s, a) = S + I + T,
\end{equation}
where the sum
\begin{equation}
S = \sum_{n=0}^{N-1}\frac1{(a + n)^s},
\end{equation}
the integral $I$ reduces to
\begin{equation}
I = \frac{(a + N)^{1-s}}{1 -s},
\end{equation}
and the tail,
\begin{equation}
T = \frac1{(a+N)^s} \left[\frac12 + \sum_{k=1}^M \frac{B_{2k}}{2k!} \frac{\langle s\rangle_n}{(a + N)^{2k-1}}\right],
\end{equation}
where the Pochammer symbol $\langle s\rangle_n$ denotes a rising factorial and we hard-code the first 51 non-zero Bernoulli numbers, $B_{2k}$. The limits $N$ and $M$, with $N \sim M$, must be chosen judiciously. Our implementation supports real $s$ and complex $a$; for a discussion of the validity of this approach, see \refcite{Johansson2015}.

We implement a polylogarithm function, defined by the sum
\begin{equation}
\text{Li}_s(a) = \sum_{n=1}^\infty \frac{a^n}{n^s},
\end{equation}
and its analytic continuation for $\Re s \le 1$ and $|a| > 1$ via our Hurwitz zeta function with the identity
\begin{equation}
\text{Li}_s(a) = \left(\frac{i}{2 \pi }\right)^{1-s} \!\! \Gamma (1-s) \left[\zeta \left(1-s,1-\frac{i \log a}{2 \pi }\right)-i^{2 s} \zeta \left(1-s,\frac{i \log a}{2 \pi }\right)\right].
\end{equation}
Our implementation supports real $s$ and complex $a$. If $s > 1$ and $|a| \le 1$, the sum is convergent and we perform direct summation of a finite number of terms. The polylogarithm functions are related to complete Fermi-Dirac integrals and Debye functions. In fact, we require in \eq{eq:poly}
\begin{equation}
\text{Li}_{5/2}(-e^{-x}) = - F_{3/2}(x),
\end{equation}
where $F_j(x)$ is a Fermi-Dirac function. The $F_{3/2}(x)$ Fermi-Dirac function is implemented in \code{GSL}.

\subsection{Taylor expansion (\normalfont\code{taylor})}

The thermal functions may be Taylor expanded about $y=0$ (see e.g., \citeeq{2.18}{Wainwright:2013maa}),
\begin{align}
\begin{split}
J_\text{B}(y^2) ={}& -\frac{\pi^4}{45} + \frac{\pi^2}{12} y^2 - \frac{\pi}{6} |\Re(y^3)| - \frac{1}{32}y^4 \log\frac{|y^2|}{a_b} \\
& -2\pi^{7/2} \sum_{n=1}^{\infty} (-1)^n \frac{\zeta(2n +1)}{(n+2)!} 
\Gamma(n + 1/2) 
\left(\frac{y^2}{4\pi^2}\right)^{n+2},
\end{split}
\end{align}
and (see e.g., \citeeq{2.19}{Wainwright:2013maa})
\begin{align}
\begin{split}
J_\text{F}(y^2) ={}& \frac{7\pi^4}{360} - \frac{\pi^2}{24} y^2 + \frac{1}{32}y^4 \log\frac{|y^2|}{a_f} \\
& -\frac14\pi^{7/2} \sum_{n=1}^{\infty} (-1)^n \frac{\zeta(2n +1)}{(n+2)!}
\Gamma(n + 1/2) 
\left(\frac{y^2}{\pi^2}\right)^{n+2} 
\frac{2^{2n+1} - 1}{2^{2n+1}}.
\end{split}
\end{align}
Note well that \citeeq{2.16}{Wainwright:2013maa} defines $J_\text{F}$ with an extra minus sign relative to our \eq{eq:def}, which we removed from \citeeq{2.19}{Wainwright:2013maa}. These expansions are implemented in \code{taylor}. 

\subsection{Derivatives}\label{sec:derivatives}

We implemented first- and second-derivatives with respect to $y^2$ of the thermal functions by taking analytic derivatives of the Bessel function representation of the thermal functions (\code{bessel}) and by numerical differentiation (\code{approx}) with \code{gsl_deriv_central}. We utilised similar relations to \eq{eq:real} to relate real parts of Bessel functions of complex arguments to Bessel functions with real arguments. Unfortunately, differentiating \eq{eq:bessel} worsens the convergence of the sum of Bessel functions, partly since $\df{f(ny)}{y}$ introduces additional factors of $n$ in the sum, and partly since differentiation lowers the index of the Bessel functions. 

\section{Results and performance}

\begin{figure}
    \centering
    \includegraphics[width=0.32\linewidth]{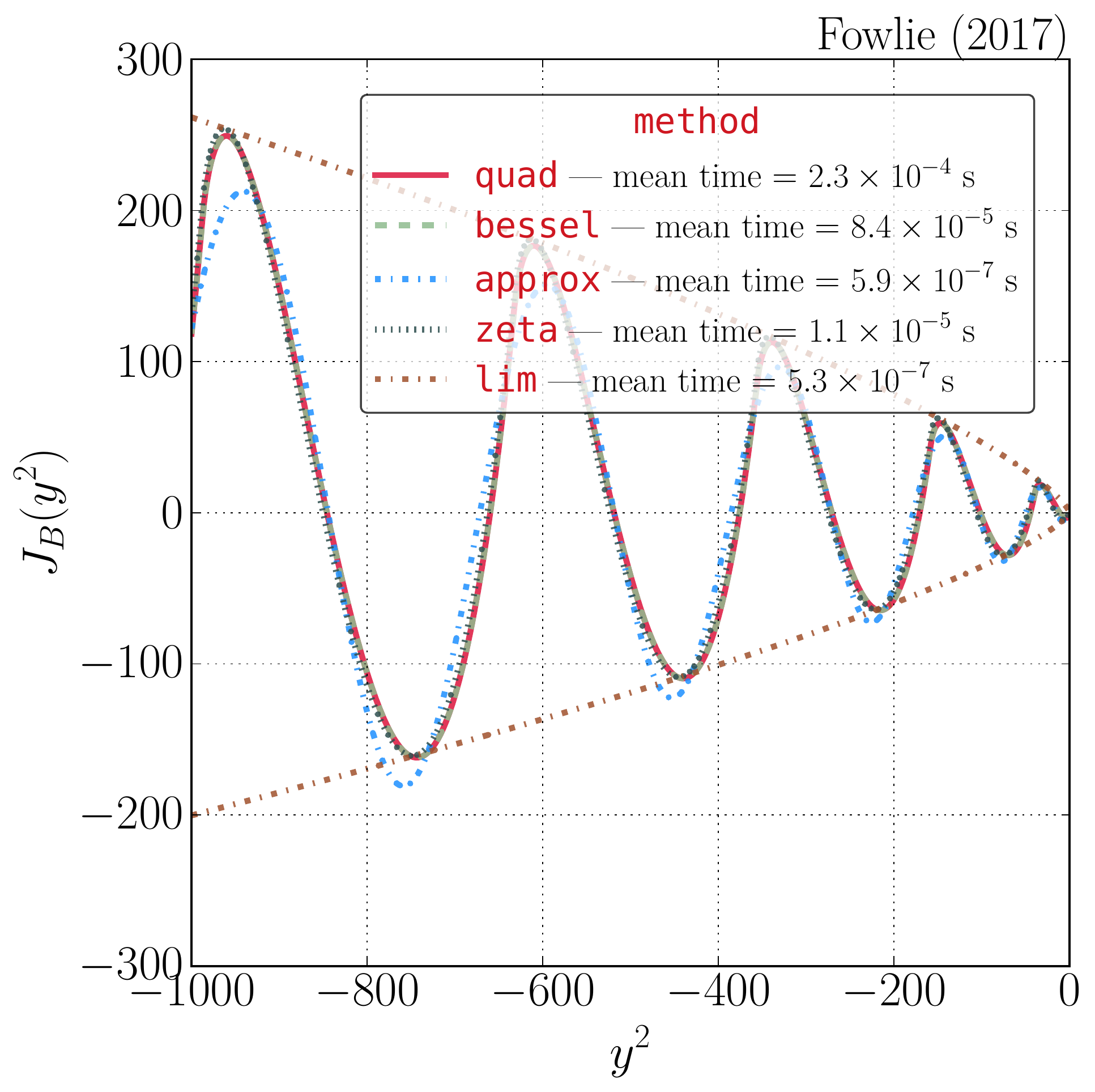}
    \includegraphics[width=0.32\linewidth]{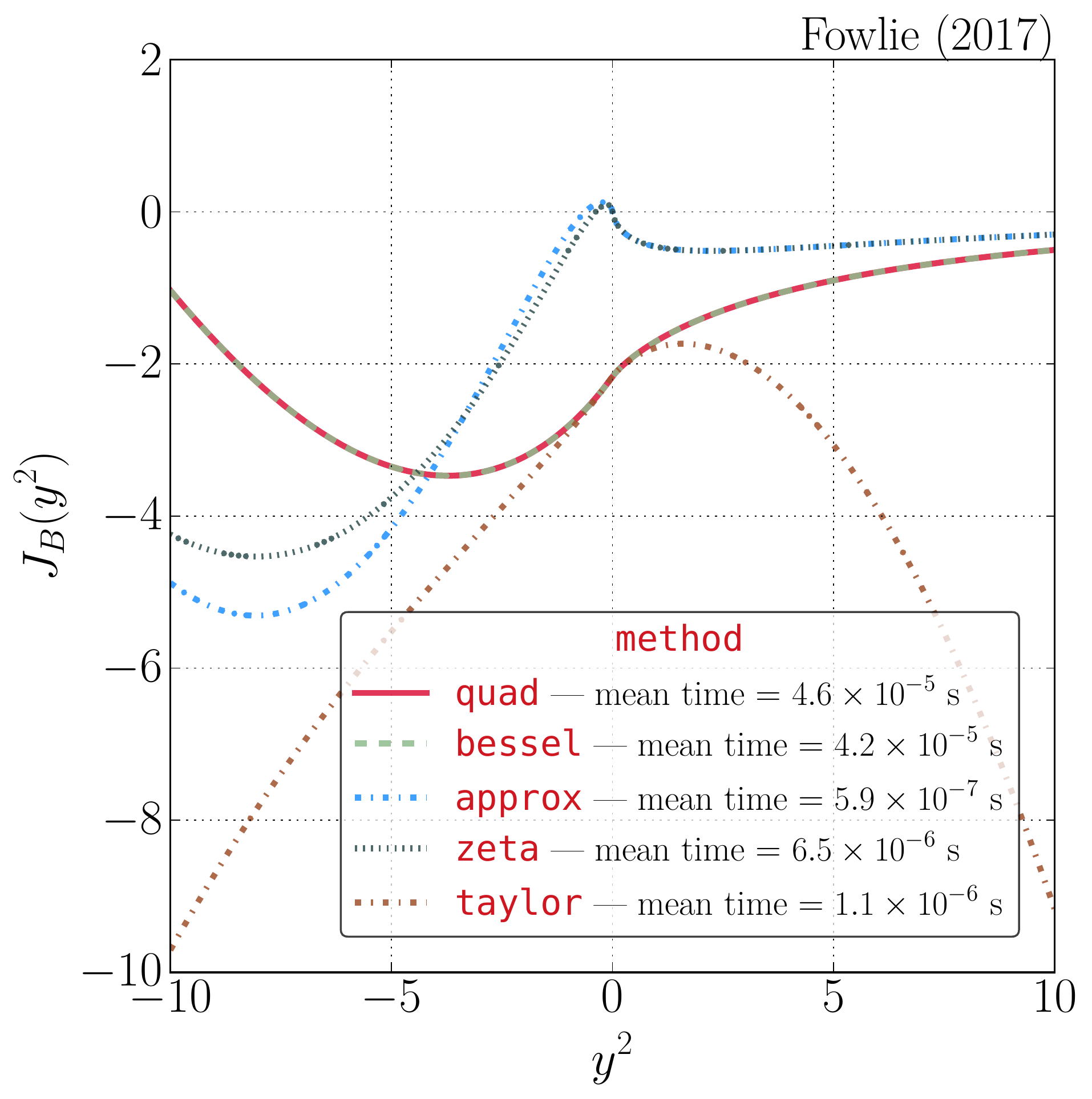}
    \includegraphics[width=0.32\linewidth]{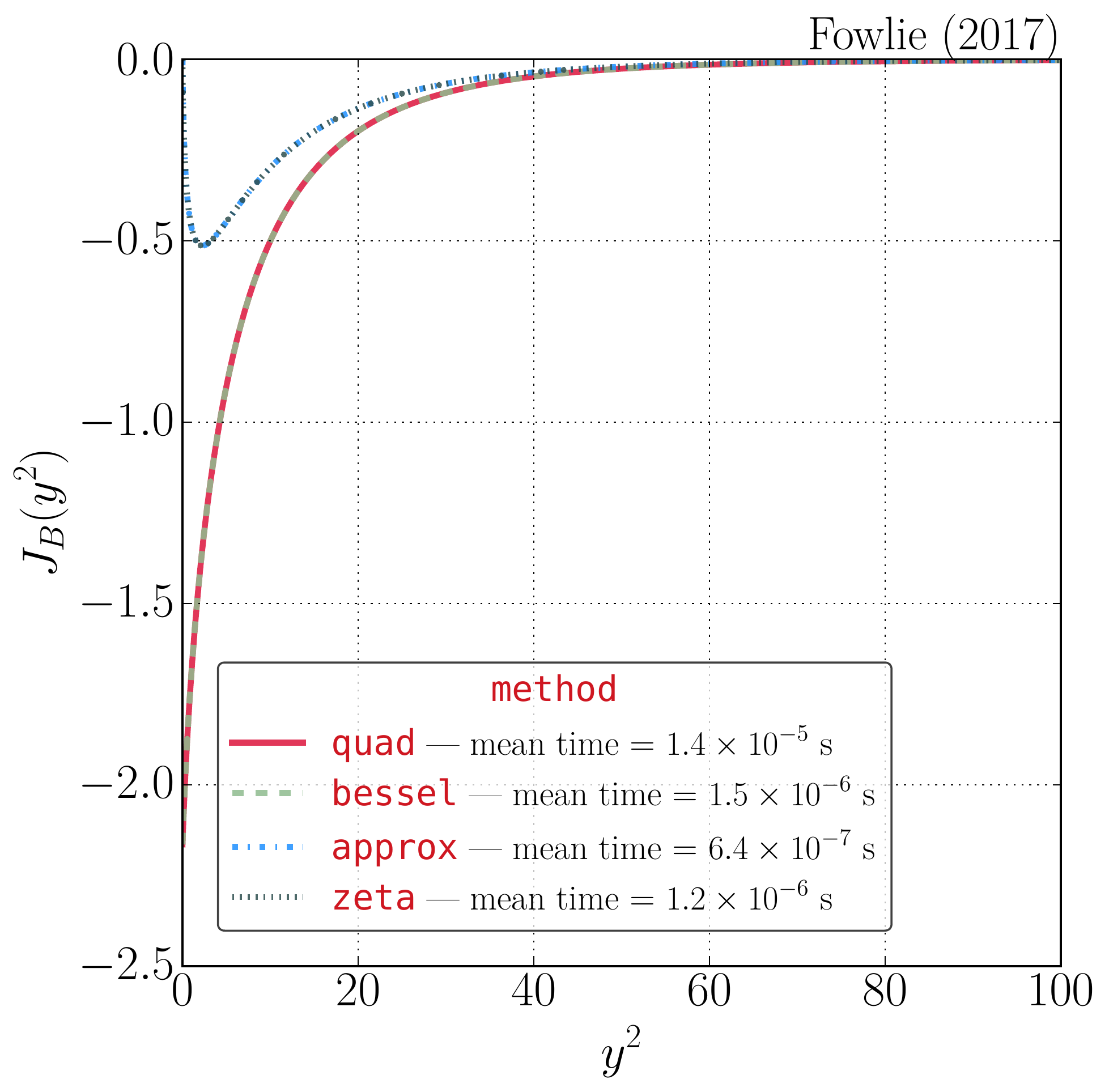}
    \caption{Thermal function $J_\text{B}(y^2)$, evaluated with \texttt{J\_B\_\{\color{fireenginered}{method}\}} for the the methods indicated in the legend and described in the text. The average time per function call is shown in the legend.}
    \label{fig:JB}
\end{figure}

\begin{figure}
    \centering
    \includegraphics[width=0.32\linewidth]{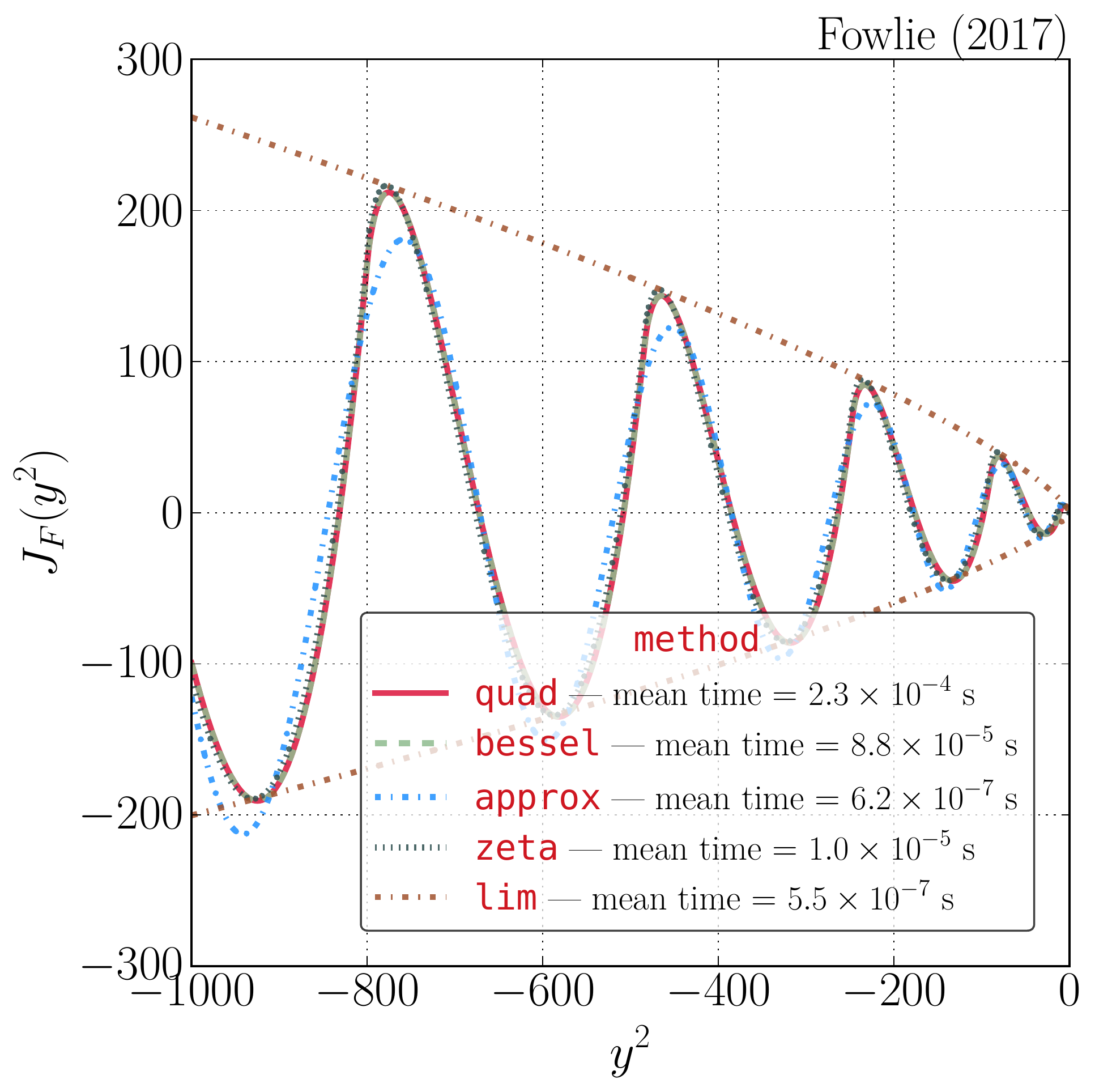}
    \includegraphics[width=0.32\linewidth]{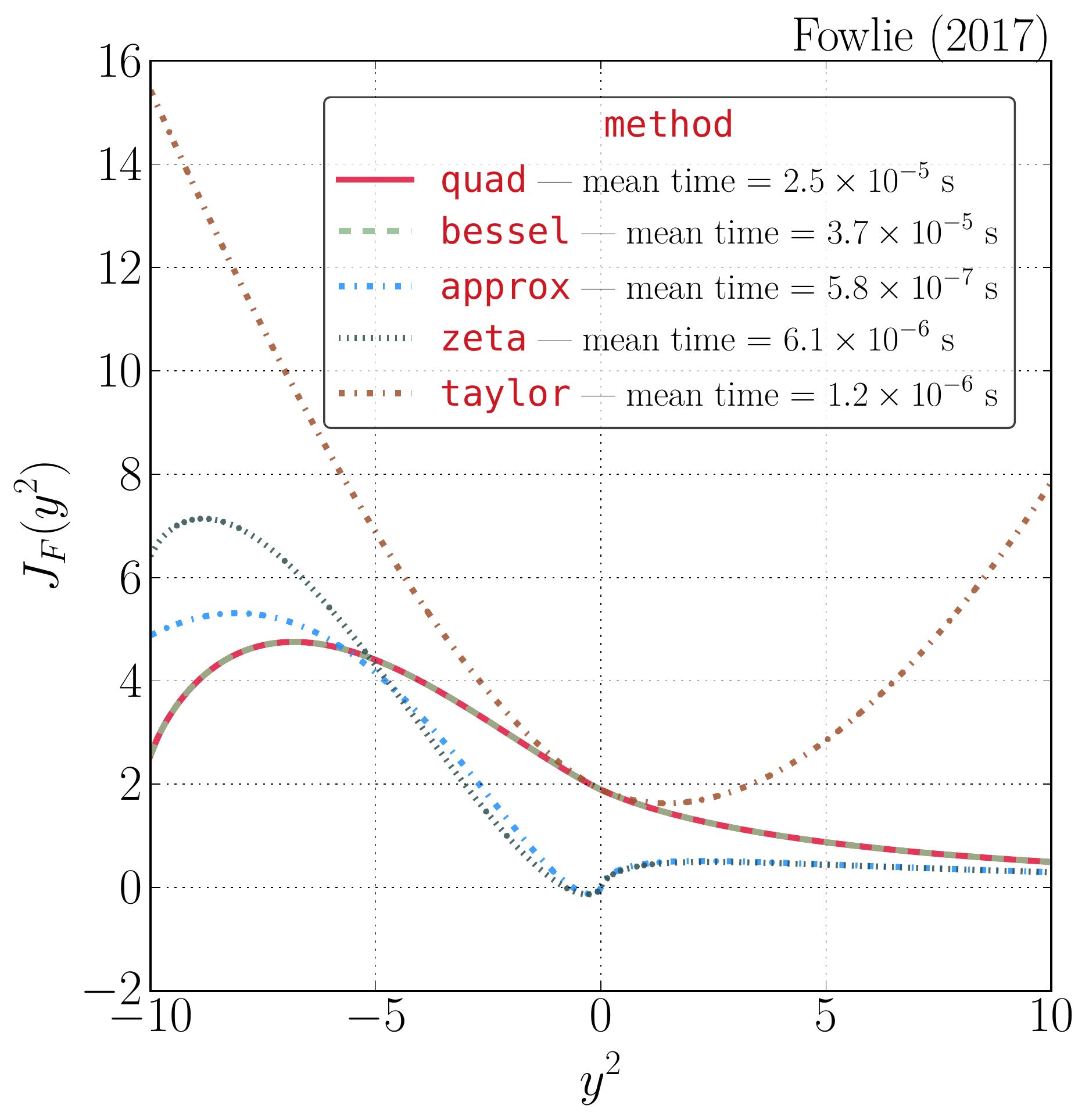}
    \includegraphics[width=0.32\linewidth]{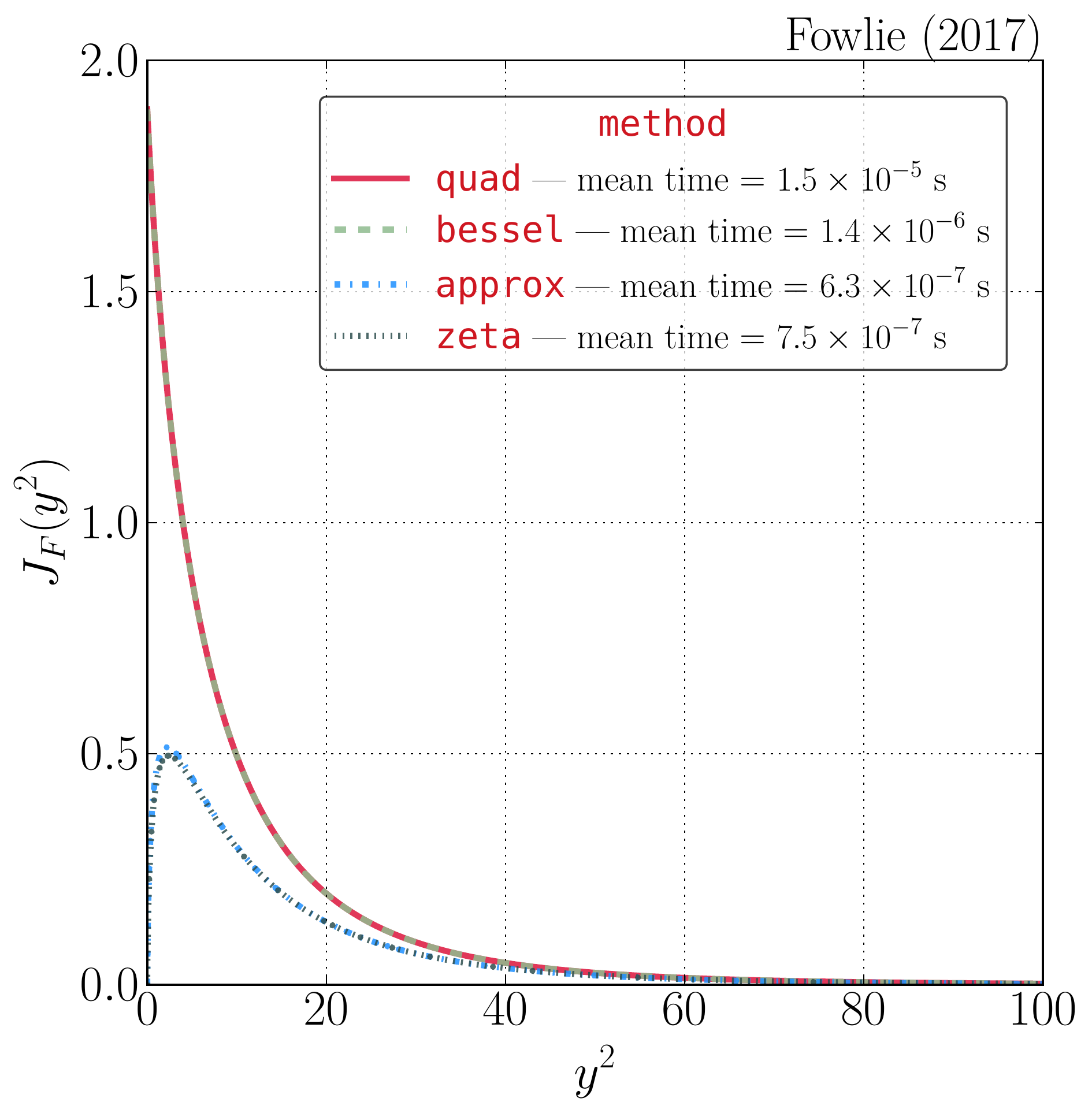}
    \caption{Thermal function $J_\text{F}(y^2)$, evaluated with \texttt{J\_F\_\{\color{fireenginered}{method}\}} for the the methods indicated in the legend and described in the text. The average time per function call is shown in the legend.}
    \label{fig:JF}
\end{figure}

\begin{figure}
    \centering
    \includegraphics[width=0.49\linewidth]{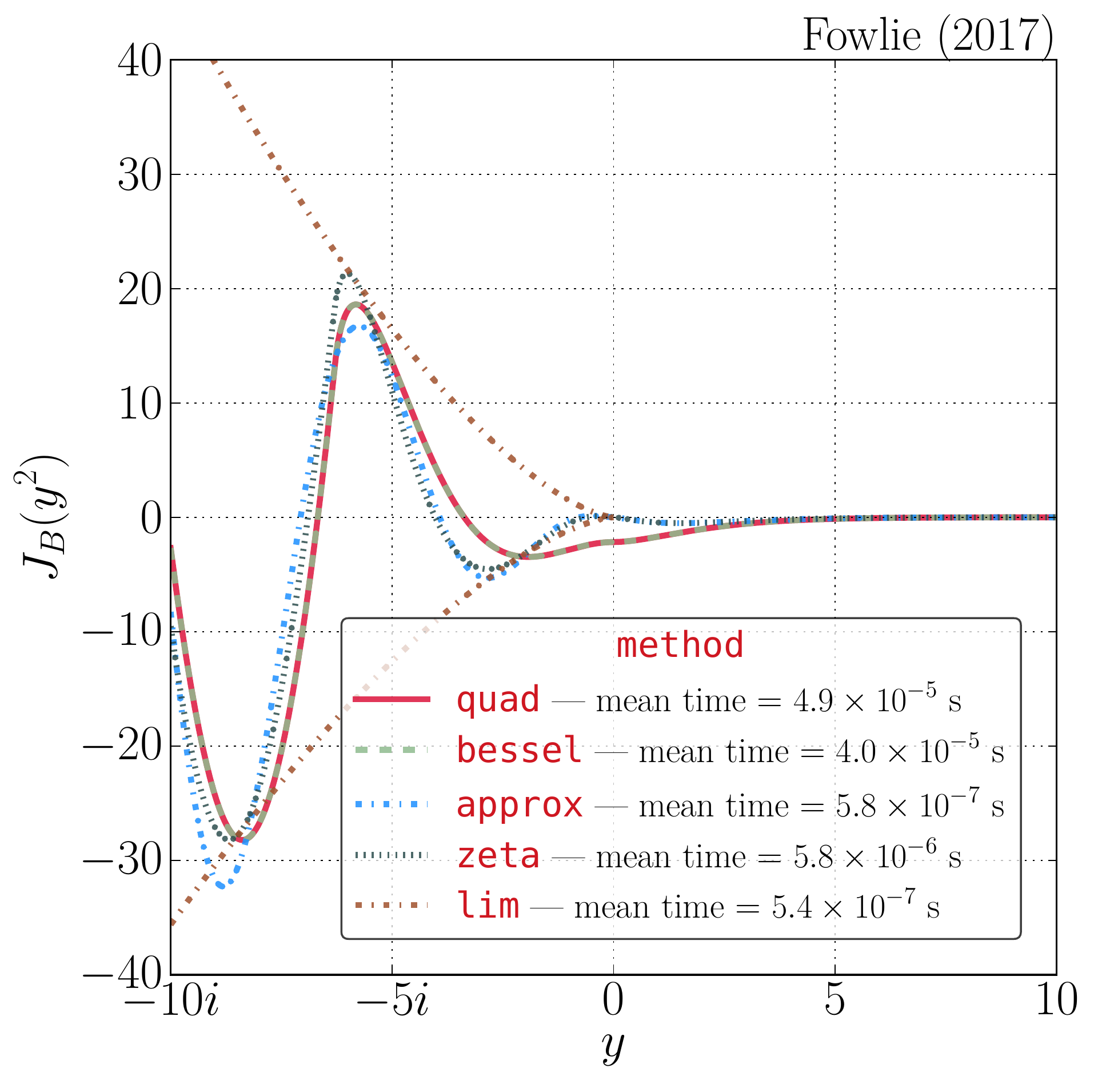}
    \includegraphics[width=0.49\linewidth]{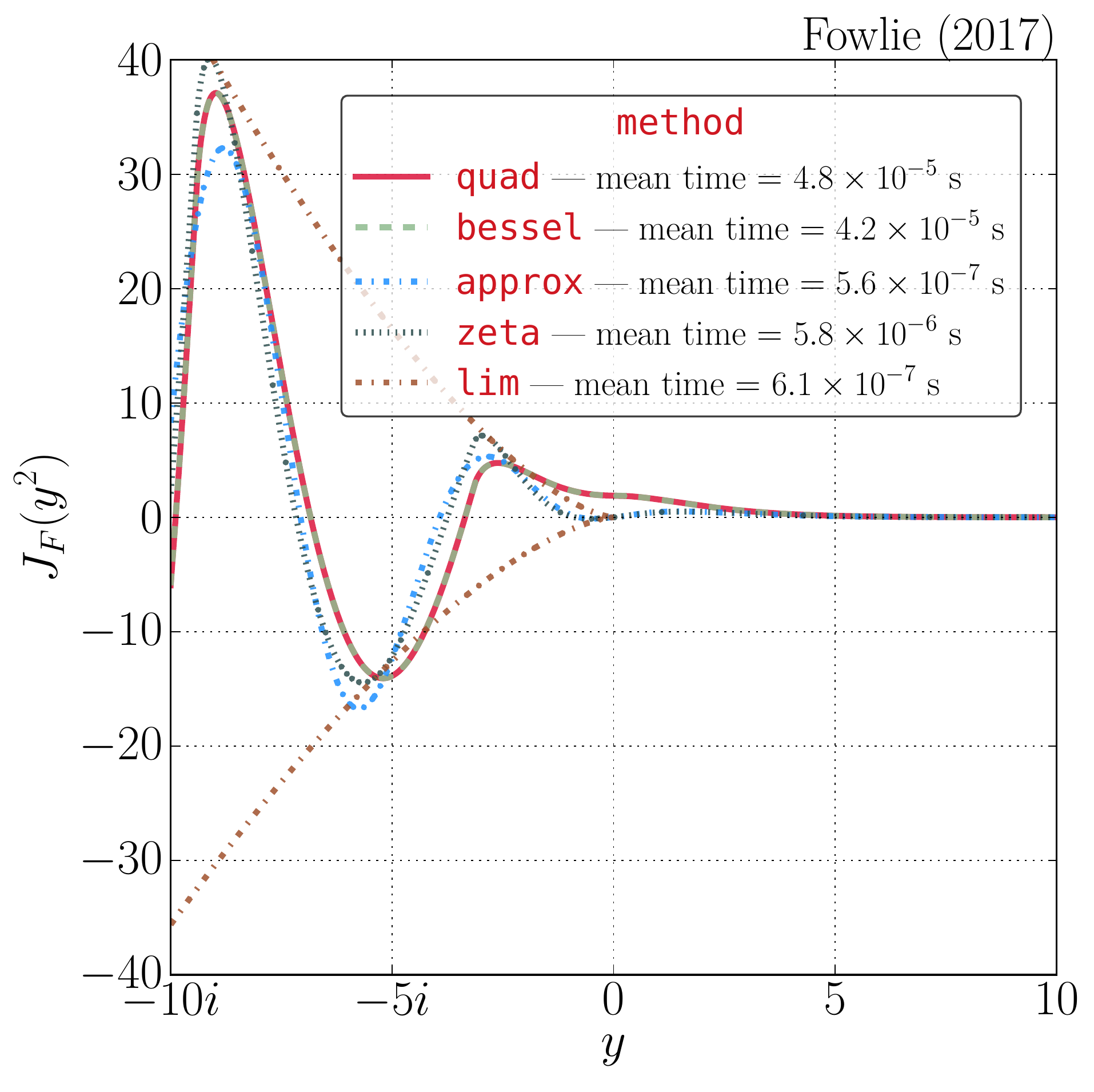}
    \caption{Thermal functions $J_\text{B/F}(y^2)$, evaluated with \texttt{J\_F\_\{\color{fireenginered}{method}\}} for the the methods indicated in the legend and described in the text. The average time per function call is shown in the legend. Note the unusual $x$-axis, which is imaginary on the left-hand side and real on the right-hand side of zero.}
    \label{fig:J}
\end{figure}

The thermal functions $J_\text{B}(y^2)$ and $J_\text{F}(y^2)$ obtained from the methods described in \refsec{sec:methods} are shown in \reffig{fig:JB} and \reffig{fig:JF}, respectively. We show them as functions of $y$ in \reffig{fig:J}. We find superb agreement between quadrature (\code{quad}) and Bessel summation (\code{bessel}) calculations of the thermal functions for all $|y^2| \lesssim 10^5$. As expected, the Taylor functions (\code{taylor}) are reasonable only inside a limited radius $|y^2| \lesssim 1$. The zeta function asymptotic representations (\code{zeta}) are fast and accurate, provided that $y^2 \gtrsim 30$ or $y^2 \lesssim -100$. For $y^2 < 0$, the leading-order approximation is about 10 times faster than the zeta function, though visibly poorer in accuracy. For $y^2 > 0$, the leading-order approximation is just as good as the zeta function representation and slightly faster. For $y^2 < 0$, the function exhibits periodicity in $y \to y + 2 n \pi i$ and saturates the upper and lower limits (\code{lim}).

We estimate that our functions completely breakdown for $y^2 \ll 0$ once the bound in \eq{eq:bound} is exceeded. For the method \code{zeta}, we reach machine limits $y^2 \sim -10^{300}$ without breaking the bound. The method \code{bessel} fails at $y^2 \sim -10^{30}$; this probably corresponds to a breakdown at $|y| \sim 10^{15}$ since a C++ \code{double} contains about 15 decimal places. The methods \code{quad} fails at about $y^2 \sim - 10^8$. By increasing the memory available in the integration (\code{gsl_integration_workspace_alloc}) via our keyword argument \code{max_n}, agreement survives until $y^2 \sim - 10^9$, but the integration becomes rather slow. This suggests that quadrature breaks down at about $y^2 \sim - 10^8$.

The average time for a single function call with $10^{-7}$ relative error and $10^{-7}$ absolute error is shown in the legends of \reffig{fig:JB} and \reffig{fig:JF}. The time depends on the argument $y^2$; the methods are slower for $y^2 < 0$. The functions typically take about $10^{-6}$ s for $y^2 \ge 0$ and slightly more for $y^2 <0$. The precision can be traded for speed, though. The Bessel function method is fastest when $y^2 \gg 0$ since fewer terms must be summed for convergence. When called within Python, the C++ thermal functions are about 10 times faster than those present in \code{CosmoTransitions}\cite{Wainwright:2011kj},
\begin{lstlisting}[language=Python]
from timeit import timeit
timeit('J_B(100.)', setup="from thermal_funcs import J_B")
timeit('Jb(10.)', setup="from finiteT import Jb")
\end{lstlisting}
When called from Mathematica, the C++ thermal functions are about 100 times faster than a naive application of \code{NIntegrate} in Mathematica. Compare e.g.,
\begin{lstlisting}[language=Mathematica]
JB[100] // RepeatedTiming
integrand = x^2 Log[1-Exp[-Sqrt[x^2+100.]]]
NIntegrate[integrand, {x, 0, \[Infinity]}] // RepeatedTiming
\end{lstlisting}
The timings were for compilation with \code{g++ 4.8.4} with \code{-03} optimisation on a $3.6$ GHz CPU. The C++ functions are significantly more accurate than a naive application of \code{NIntegrate} and those in \code{CosmoTransitions}, especially for $y^2 < 0$. 

\begin{figure}
    \centering
    \includegraphics[width=0.48\linewidth]{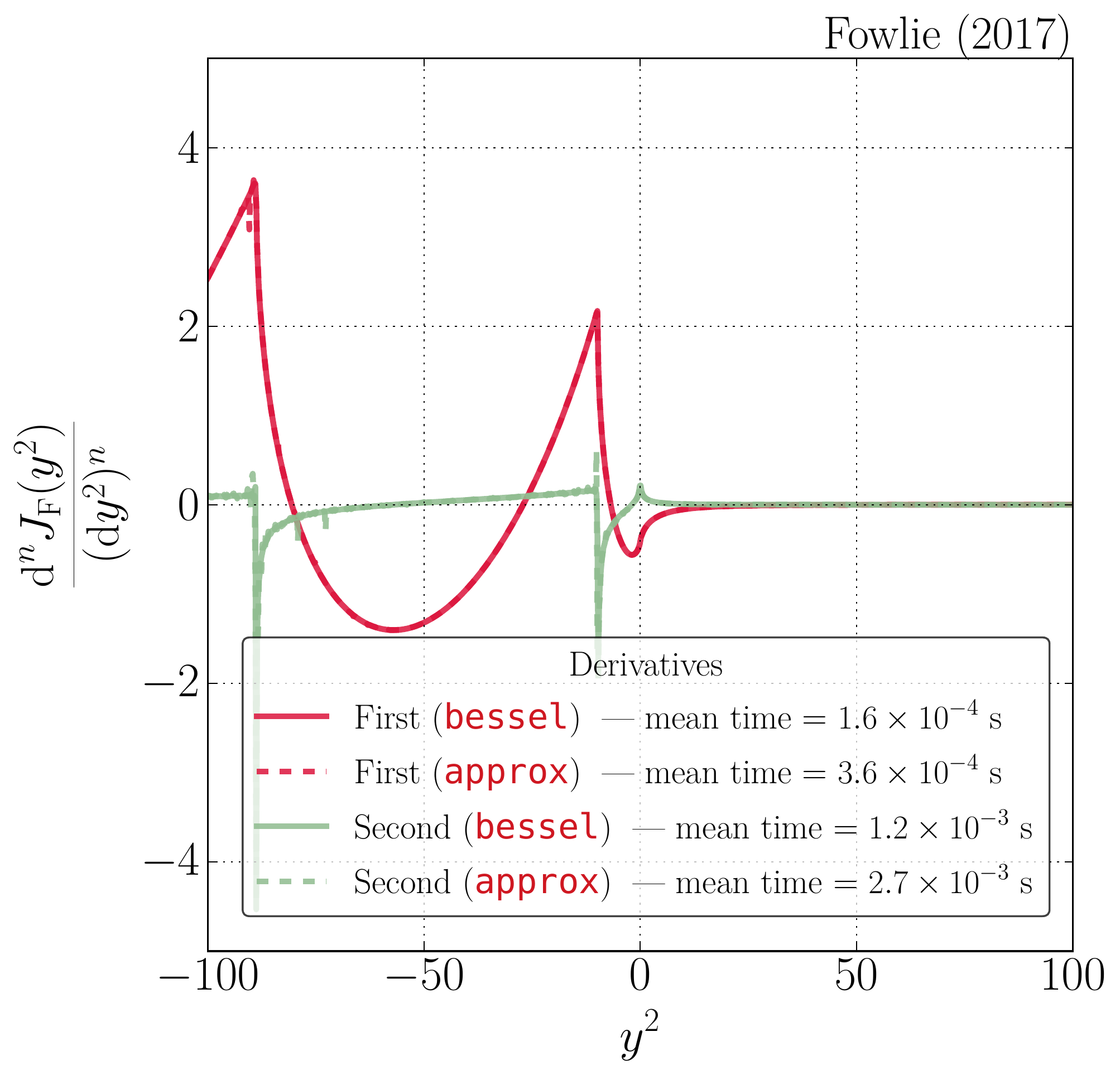}
    \includegraphics[width=0.48\linewidth]{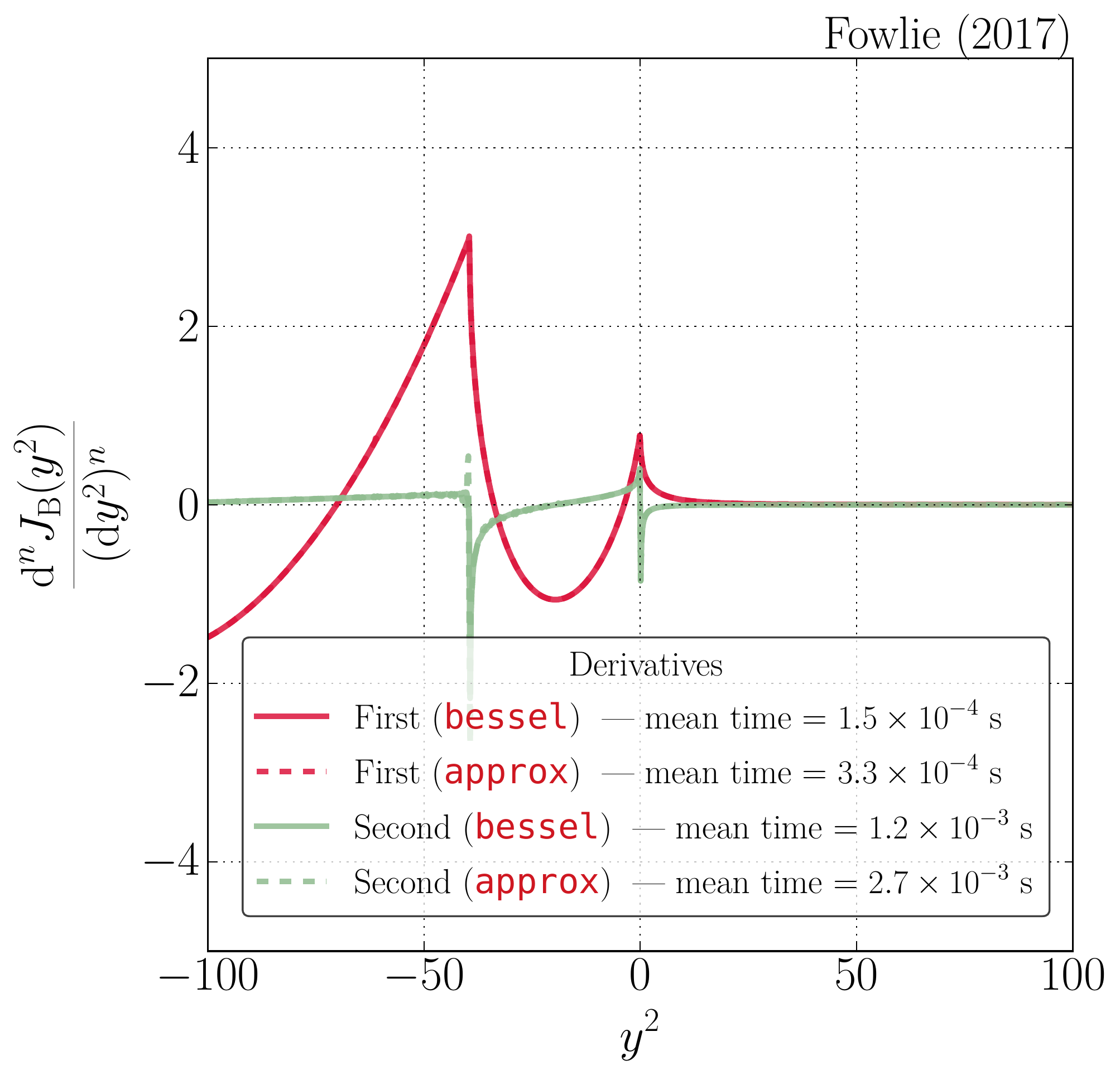}
    \caption{First and second derivatives of thermal functions $J_\text{F}(y^2)$ (left) and $J_\text{B}(y^2)$ (right) evaluated with \texttt{D\{1/2\}\_J\_\{F/B\}\_\{\color{fireenginered}{method}\}} for the the methods indicated in the legend and described in \refsec{sec:derivatives}. The average time per function call is shown in the legend.}
    \label{fig:D_J}
\end{figure}

As shown in \reffig{fig:D_J}, the analytic first- and second-derivatives (\code{bessel}) of the thermal functions were in agreement with numerical derivatives (\code{approx}). Numerical differentiation suffered from numerical noise and was slightly slower for first derivatives. As discussed, however, differentiation appears to worsen the rate of convergence of the sum of Bessel functions, such that the second derivatives with \code{bessel} require many terms before convergence and are slightly slower than numerical differentiation.

\section{Installation and API}\label{sec:api}

The code is available from \url{https://github.com/andrewfowlie/thermal_funcs}. The library should be built via \code{make lib} in the main directory. This should build the \code{/lib/thermal_funcs.so} library. This requires a C++ compiler and the \code{GSL} libraries \code{libgsl.so} and \code{libgslcblas.so}, which may be installed in Ubuntu by e.g., \code{apt-get install libgsl0ldbl}.\footnote{See \url{https://www.gnu.org/software/gsl/}.} The code was tested with the C++ compilers \code{g++} and \code{icc}, and uses the former by default.

The header file \code{/src/thermal_funcs.h} is shown in \reffig{fig:thermal_funcs.h}. Macros in the header file enable C++ and C compatibility (which is required for a Mathematica interface). The header file declares twelve functions with return type \code{double} named \code|J_{B/F}_{method}|, where method may be one of \code{quad}, \code{bessel}, \code{taylor}, \code{lim}, \code{approx} or \code{zeta}, e.g., \code{J_F_quad}. The functions have a single mandatory argument: \code{double y_squared}. 

\begin{figure}[ht]
    \centering
\begin{lstlisting}[linerange={20-51}]
double J_B_quad(double y_squared,
                double abs_error DEFAULT(1E-7),
                double rel_error DEFAULT(1E-7),
                int max_n DEFAULT(10000));
double J_F_quad(double y_squared,
                double abs_error DEFAULT(1E-7),
                double rel_error DEFAULT(1E-7),
                int max_n DEFAULT(10000));
double J_B_taylor(double y_squared,
                  double abs_error DEFAULT(1E-7),
                  double rel_error DEFAULT(1E-7),
                  int max_n DEFAULT(10000));
double J_F_taylor(double y_squared,
                  double abs_error DEFAULT(1E-7),
                  double rel_error DEFAULT(1E-7),
                  int max_n DEFAULT(10000));
double J_F_bessel(double y_squared,
                  double abs_error DEFAULT(1E-7),
                  double rel_error DEFAULT(1E-7),
                  int max_n DEFAULT(10000),
                  bool fast DEFAULT(false));
double J_B_bessel(double y_squared,
                  double abs_error DEFAULT(1E-7),
                  double rel_error DEFAULT(1E-7),
                  int max_n DEFAULT(10000),
                  bool fast DEFAULT(false));
double J_B_lim(double y_squared, bool upper DEFAULT(true));
double J_F_lim(double y_squared, bool upper DEFAULT(true));
double J_B_approx(double y_squared);
double J_F_approx(double y_squared);
double J_B_zeta(double y_squared, int n DEFAULT(25));
double J_F_zeta(double y_squared, int n DEFAULT(25));
\end{lstlisting}
    \caption{Function declarations in \code{thermal_funcs.h} header file.}
    \label{fig:thermal_funcs.h}
\end{figure}

The methods \code{quad}, \code{bessel} and \code{taylor} accept additional arguments: 
\begin{itemize}
\item \code{double abs_error DEFAULT(1E-7)}: The absolute error permissible in a summation or integral.
\item \code{double rel_error DEFAULT(1E-7)}: The relative error permissible in a summation or integral. 
\item \code{int max_n DEFAULT(10000)}: The maximum number of terms in a sum or the maximum number of subdivisions for numerical integration.
\end{itemize}
The method \code{bessel} accepts a further argument, \code{bool fast DEFAULT(false)}, governing whether asymptotic formulae are used for Bessel functions. The method \code{zeta} accepts a single optional argument, \code{int n DEFAULT(25)}, governing the number of terms in the Euler-Maclaurin summation of the Hurwitz zeta function. The method \code{lim} accepts a single optional argument, \code{bool upper DEFAULT(true)}, which selects the upper or lower limit. The functions \code{J_F_lim} and \code{J_B_lim} are identical; we provide both to maintain consistency.

\begin{sloppypar}
The first and second derivatives of the thermal functions are declared in \code{/src/derivatives.h}, which is shown in \reffig{fig:derivatives.h}. The header file declares eight functions with return type \code{double} named \code|D{1/2}_J_{B/F}_{bessel/approx}|, e.g., \code{D1_J_F_bessel}, and a single mandatory argument, \code{double y_squared}.
The functions \code|D{1/2}_J_F_approx| accepts four optional arguments: \code{double step DEFAULT(1E-1)}, that specifies the step size in the numerical differentiation, and \code{double abs_error DEFAULT(1E-7)}, \code{double rel_error DEFAULT(1E-7)} and \code{int max_n DEFAULT(10000)}, as described above.
The functions \code|D{1/2}_J_F_bessel| accept the latter three optional arguments.
The Hurwitz zeta function and polylogarithm function in \code{/lib/thermal_funcs.so} are declared in the header file \code{/src/zeta.h}. 
\end{sloppypar}
\begin{figure}[ht]
    \centering
\begin{lstlisting}[linerange={25-63}]
double D1_J_F_bessel(double y_squared,
                     double abs_error DEFAULT(1E-7),
                     double rel_error DEFAULT(1E-7),
                     int max_n DEFAULT(10000));
double D1_J_B_bessel(double y_squared,
                     double abs_error DEFAULT(1E-7),
                     double rel_error DEFAULT(1E-7),
                     int max_n DEFAULT(10000));

double D2_J_F_bessel(double y_squared,
                     double abs_error DEFAULT(1E-7),
                     double rel_error DEFAULT(1E-7),
                     int max_n DEFAULT(10000));
double D2_J_B_bessel(double y_squared,
                     double abs_error DEFAULT(1E-7),
                     double rel_error DEFAULT(1E-7),
                     int max_n DEFAULT(10000));

double D1_J_F_approx(double y_squared,
                     double step DEFAULT(1E-2),
                     double abs_error DEFAULT(1E-7),
                     double rel_error DEFAULT(1E-7),
                     int max_n DEFAULT(10000));
double D1_J_B_approx(double y_squared,
                     double step DEFAULT(1E-2),
                     double abs_error DEFAULT(1E-7),
                     double rel_error DEFAULT(1E-7),
                     int max_n DEFAULT(10000));

double D2_J_F_approx(double y_squared,
                     double step DEFAULT(1E-1),
                     double abs_error DEFAULT(1E-7),
                     double rel_error DEFAULT(1E-7),
                     int max_n DEFAULT(10000));
double D2_J_B_approx(double y_squared,
                     double step DEFAULT(1E-1),
                     double abs_error DEFAULT(1E-7),
                     double rel_error DEFAULT(1E-7),
                     int max_n DEFAULT(10000));
\end{lstlisting}
    \caption{Function declarations in \code{derivatives.h} header file.}
    \label{fig:derivatives.h}
\end{figure}

There is an example program \code{/src/example.cpp} shown in \reffig{fig:example.cpp}. The command \code{make example} builds this code into a small program, \code{./bin/example}, that prints an evaluation of a thermal function and its derivatives.

\begin{figure}[ht]
    \centering
\begin{lstlisting}[linerange={9-19}]
#include <stdio.h>
#include <thermal_funcs.h>
#include <derivatives.h>


int main() {
    printf("J_B = %e\n", J_B_bessel(100.));
    printf("D1_J_B_bessel = %e\n", D1_J_B_bessel(100.));
    printf("D2_J_B_bessel = %e\n", D2_J_B_bessel(100.));
    return 0;
}
\end{lstlisting}
    \caption{Example program, \code{example.cpp}, built by \code{make example}.}
    \label{fig:example.cpp}
\end{figure}

\subsection{Python}

The command \code{make python} should build the Python interface via \code{SWIG}. This requires \code{SWIG}\footnote{See \url{http://www.swig.org/}.}, which may be installed via e.g., \code{apt-get install swig} in Ubuntu. The interface requires the \code{Python.h} header file, provided with e.g., 
\texttt{apt-get install python-dev}
in Ubuntu. The header file is located automatically by \code{pkg-config}; if that fails, set the \code{PYTHON} variable manually in \code{/src/makefile} to the location of the \code{Python.h} header file. The interface may be built for Python 2 or 3, so long as it built with the appropriate \code{Python.h} header file.

Once built, from a Python session within the  main directory (or from within any directory if \code{thermal_funcs} is added to your \code{PYTHONPATH}), import the functions with
\begin{lstlisting}[language=Python]  
from thermal_funcs import J_B, J_F
\end{lstlisting}
The functions accept a single mandatory argument, $y^2$. The optional keyword argument \code{method = 'bessel'}, which should be \code{'quad'}, \code{'bessel'}, \code{'taylor'}, \code{'lim'}, \code{'approx'} or \code{'zeta'}, governs which method to employ. The C++ optional arguments in \refsec{sec:api} are keyword arguments that are passed if applicable. E.g.
\begin{lstlisting}[language=Python]  
J_F(100., method='quad', rel_error=1E-2, abs_error=1E-6, max_n=1000)
\end{lstlisting}
calculates $J_F(y^2 = 100)$ with method \code{quad}, and with optional keyword arguments that specify the desired precision. First and second derivatives are calculated via the keyword argument \code{derivative = 1} or \code{derivative = 2}, respectively. If derivatives are required, \code{method} must be  \code{'approx'} or  \code{'bessel'}. E.g.,
\begin{lstlisting}[language=Python]  
J_F(100., derivative=2)
\end{lstlisting}
calculates the second derivative. The \code{'approx'} method accepts an optional argument, \code{step}, that specifies the step size in the numerical differentiation. E.g.,
\begin{lstlisting}[language=Python]  
J_F(100., method='approx', step=0.1, derivative=2)
\end{lstlisting}

\subsection{Mathematica}

The Mathematica interface should be built by \code{make mathematica}. It requires Mathematica 11 or later and the \code{uuid} library, which may be installed in Ubuntu by e.g., \code{sudo apt-get install uuid-dev}. This utilises the Mathematica \code{WSTP} interface (the new name for \code{MathLink}).
The command \code{make mathematica} should locate it automatically in Linux or a system with \code{math} in the path. If it fails, it may be necessary to set the \code{MATH_INC} variable manually in \code{/src/makefile} to the result of the Mathematica command \begin{lstlisting}[language=Mathematica] 
FileNameJoin[{$InstallationDirectory, 
              "SystemFiles", "Links", "WSTP", "DeveloperKit", 
              $SystemID, 
              "CompilerAdditions"}]
\end{lstlisting}
The Mathematica interface should be invoked by \code[language=Mathematica]{Install["./src/math.exe"]}. This provides the \code{bessel} method with default arguments, since it is fast and robust, in functions named \code{JB} and \code{JF}. E.g.,
\begin{lstlisting}[language=Mathematica]  
Install["./src/math.exe"]
Plot[{JB[ysq], JF[ysq]}, {ysq, -1000, 100}]
\end{lstlisting}
should install and plot the thermal functions. First and second derivatives are calculated by a keyword argument, which should be \code{1} or \code{2}, respectively. E.g.,
\begin{lstlisting}[language=Mathematica]  
JB[100., derivative -> 2]
\end{lstlisting}
calculates the second derivative. The symbolic derivatives are defined such that e.g.,
\begin{lstlisting}[language=Mathematica]  
D[JB[x], x]
JB'[x]
JB[x, derivative -> 1]
\end{lstlisting}
are all equivalent. The C++ optional arguments in \refsec{sec:api} are keyword arguments, though in camel case (since underscores are are forbidden). E.g.,
\begin{lstlisting}[language=Mathematica]  
JB[100., absError -> 1*^-10]
\end{lstlisting}

\section{Conclusions}

We presented a C++ implementation of thermal functions (and their first and second derivatives) that appear finite-temperature quantum field theory. The functions have no closed form and appear in no standard libraries. The functions play a role in determining the history of Universe via the free-energy of scalar fields. The code is available from \url{https://github.com/andrewfowlie/thermal_funcs}.

\section*{Acknowledgements}
This work in part was supported
by  the  ARC  Centre  of  Excellence  for  Particle  Physics
at  the  Terascale.

\bibliography{main}
\bibliographystyle{utphys}
\end{document}